\documentclass[prl,aps,twocolumn,showpacs]{revtex4}

\usepackage{bm}
\usepackage[]{graphicx}
\usepackage{amsmath}
\usepackage{amssymb}
\usepackage{color}
\usepackage{MnSymbol}
\usepackage{wasysym}

\begin{document}

\title{Oscillatory Flows Induced by Microorganisms Swimming in Two-dimensions}

\author{Jeffrey S. Guasto$^{1}$, Karl A. Johnson$^{2}$, and J.P. Gollub$^{1,3}$}
\affiliation{$^{1}$Department of Physics, Haverford College, Haverford, Pennsylvania 19041, USA}
\affiliation{$^{2}$Department of Biology, Haverford College, Haverford, Pennsylvania 19041, USA}
\affiliation{$^{3}$Department of Physics, University of Pennsylvania,\\ Philadelphia, Pennsylvania 19104, USA}

\date{\today}

\begin{abstract}
We present the first time-resolved measurements of the oscillatory velocity field induced by swimming unicellular microorganisms.
Confinement of the green alga \emph{C. reinhardtii} in stabilized thin liquid films allows simultaneous tracking of cells and tracer particles. 
The measured velocity field  reveals complex time-dependent flow structures, and scales inversely with distance. 
The instantaneous mechanical power generated by the cells is measured from the velocity fields and peaks at 15 fW.  
The dissipation per cycle is more than four times what steady swimming would require.  
\end{abstract}


\maketitle

Understanding the hydrodynamics of swimming microorganisms can give insight into complex biological processes such as predator/prey interactions and flagellar mechanics \cite{Sheng2007,Mitchison2010}.
A variety of mathematical models have been proposed to describe interactions between swimming cells and their environment \cite{Alexander2008,Underhill2008,Lauga2009}.
However, experimental studies of velocity fields are rare \cite{Drescher2010}, and oscillatory near-field hydrodynamics have not been investigated experimentally despite suggestion of their importance by some numerical studies \cite{Ishikawa2007,Pooley2007}.

The flows induced by swimming organisms are responsible for biogenic mixing \cite{Katija2009}, and microorganisms in particular enhance the transport properties of suspensions \cite{Underhill2008,Leptos2009}.
Active particles and microorganisms also affect the rheological properties of suspensions, where the effective viscosity is increased in the case of ``pullers'' (e.g. \emph{Chlamydomonas reinhardtii}) and decreased for ``pushers'' (e.g. \emph{E. coli}) \cite{Rafai2010,Sokolov2009}.
Hydrodynamic interactions leading to coherent motion have been studied for concentrated bacterial cells \cite{Wu2000}, but not for algal cells.

In many of these phenomena, the oscillatory swimming motion of organisms may be important.
Furthermore, the concentration of swimmers may be sufficiently high that organisms are within several radii of one another, close enough that near-field effects should be considered.
Understanding the resulting time-dependent flow fields is required for the interpretation of interactions between swimming cells and their environments.

We present the first time-resolved measurements of the flow field around oscillatory swimming microorganisms (\emph{C. reinhardtii}).
This is accomplished through simultaneous tracking of swimming cells and passive tracer particles within quasi-two-dimensional (quasi-2D) environments ($\approx 15$ $\mu$m thick liquid films) \cite{Prasad2009}, where direct optical access and long observation times are achieved.
These measurements reveal complex near-field flow structures that evolve throughout the beat cycle, and provide a richer picture than time-averaged flow fields and simplified models.
The instantaneous mechanical power transferred by flagella to the fluid is measured via the viscous dissipation.
These observations carry important implications for the interpretation and modeling of transport processes, locomotion, and flagellar mechanics.

\emph{C. reinhardtii} is a single-celled swimming alga with a 7-10 $\mu$m diameter body \cite{Harris2009}.
It typifies many biflagellated microorganisms with two 10-12 $\mu$m long flagella located at the anterior of the organisms.
The flagella beat at 50-60 Hz propelling the cell body in an oscillatory manner (see video \cite{EPAPS}) with a mean velocity of 100-200 $\mu$m/s.
The flagella are actuated along their length by dynein motors and take on asymmetric conformations during the power and recovery strokes to overcome the time reversibility of low Reynolds number flows.

Wild-type \emph{C. reinhardtii} (cc1690) is grown in minimal media (M1) on a light cycle (16 hr bright/8 hr dark) to ensure uniform size distribution and motility \cite{Harris2009}. 
An adjustable wire-frame device \cite{Sokolov2009} is used to stretch a 2 $\mu$l droplet of the cell suspension into a square film with side length $L=6$ mm and measured thickness $h=15 \pm 2$ $\mu$m.
This thickness is sufficient to achieve 2D hydrodynamics for cells with radius $R \approx 3.5$ $\mu$m \cite{Prasad2009}.
A trace amount of gentle surfactant (Tween 20) stabilizes the film without harming the cells.
Polystyrene microspheres (1 $\mu$m diam., Thermo Scientific) are passivated with BSA to prevent adhesion to flagella \cite{Weibel2005} and incorporated with the cells prior to stretching the film.

The device is mounted on an upright microscope where the suspensions are observed with a $40\times$ 0.75 NA objective under bright field illumination with red light ($>610$ nm) to prevent phototaxis \cite{Teplitski2004}.
A high-speed digital camera captures the motion over about 3000 frames at 50 or 500 fps.
The cells and tracer particles are segmented from one another based on size and tracked in time using a predictive algorithm \cite{Ouellette2006}.

\begin{figure}[t]
	\begin{center}
	\includegraphics[clip=true,keepaspectratio,width=7cm]{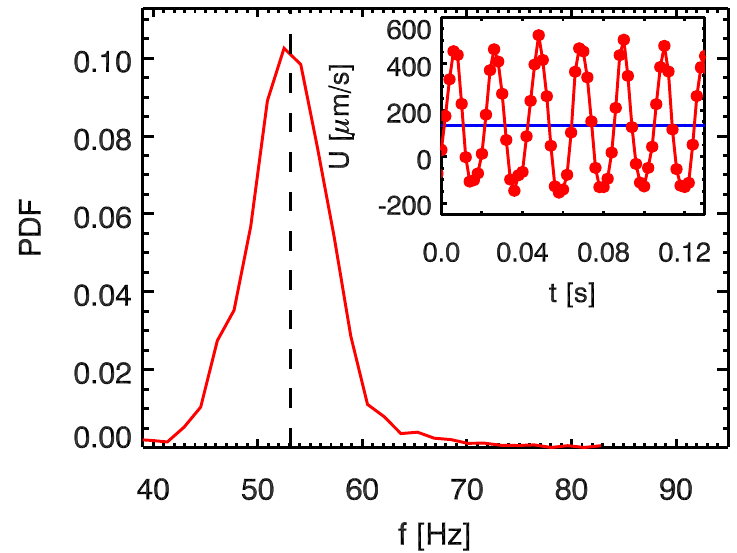}
	\end{center}
	\caption{(color online) Probability density function (PDF) of flagellar beat frequencies, $f$, measured from unicellular alga (\emph{C. reinhardtii}) swimming in quasi-2D liquid films ($\bar{f}=53 \pm 5$ Hz). Inset: velocity oscillations of a single swimming cell measured at 500 fps (red \textcolor{red}{\CIRCLE}), where the maximum velocity is four times the mean value (solid blue line, 134 $\mu$m/s).}
	\label{fig:1}
\end{figure}

The thin liquid film ensures that the algae are coincident with the focal plane of the objective and prevents distortion from out-of-focus particles. 
This allows detailed observation of the cells (mean speed $U_0=134$ $\mu$m/s) for up to about 8 s at 50 fps. 
Using high-speed imaging (500 fps), oscillations of the cell body, $U(t)$, become evident [Fig. \ref{fig:1} (inset)].
The peak forward velocity is four times the mean value and can be negative during the recovery stroke.
The probability density function (PDF) of the beat frequency $f$ for $\approx 100$ organisms is shown in Fig. \ref{fig:1}.
The oscillations have a fairly narrow distribution with a mean frequency $\bar{f}=53 \pm 5$ Hz corresponding to a beat cycle period $T = 1/\bar{f} = 18.9$ ms.

\begin{figure}[t]
\begin{center}
\includegraphics[clip=true,keepaspectratio,width=7cm]{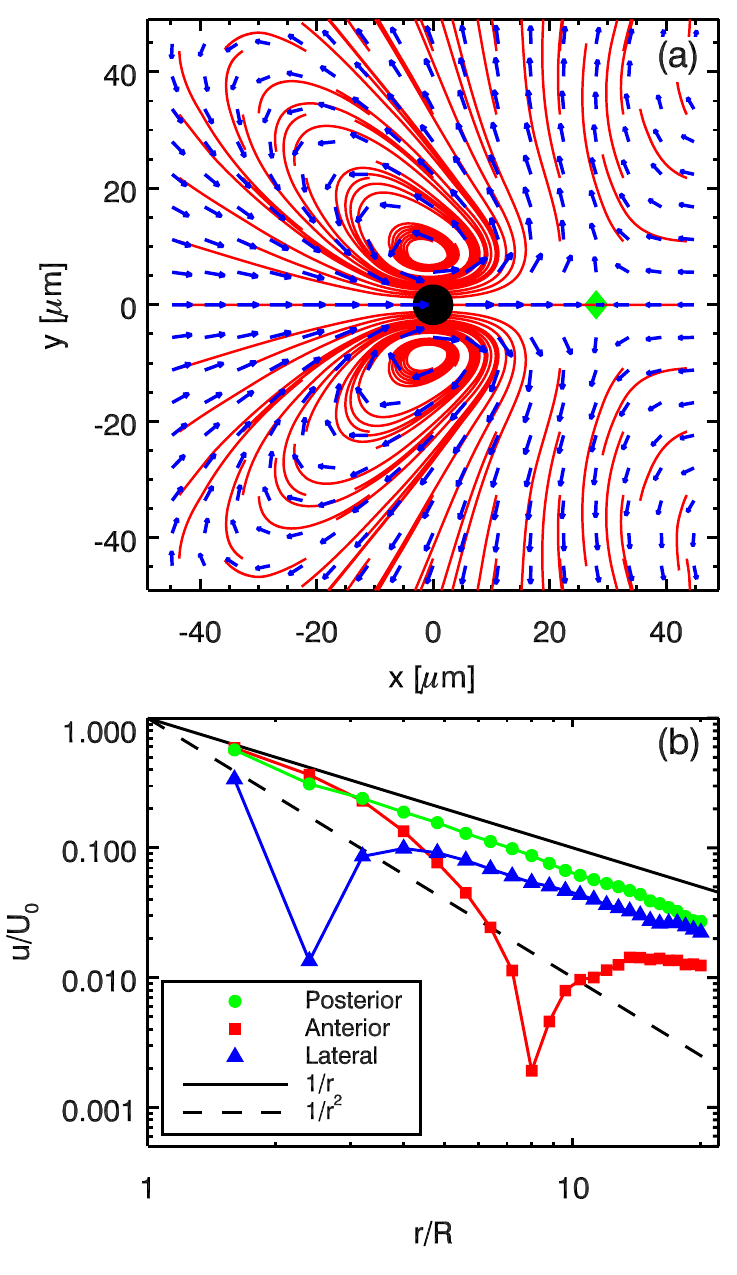}
\end{center}
\caption{(color online) (a) The time-averaged velocity field around a swimming \emph{C. reinhardtii} (black disc) in the lab frame, where the direction of travel is to the right toward the hyperbolic stagnation point (green {\Large \textcolor{green}{$\filleddiamond$}}). Solid (red) lines are instantaneous streamlines and velocity vectors are shown on a log scale \cite{JSGLogScaling}. (b) The fluid velocity magnitude in various directions away from the cell demonstrates the predicted $u \sim r^{-1}$ scaling for a force dipole in 2D. Local minima correspond to stagnation points encountered in some directions.}
\label{fig:2}
\end{figure}

\begin{figure*}[t]
\begin{center}
\includegraphics*[clip=true,keepaspectratio,width=15.95cm]{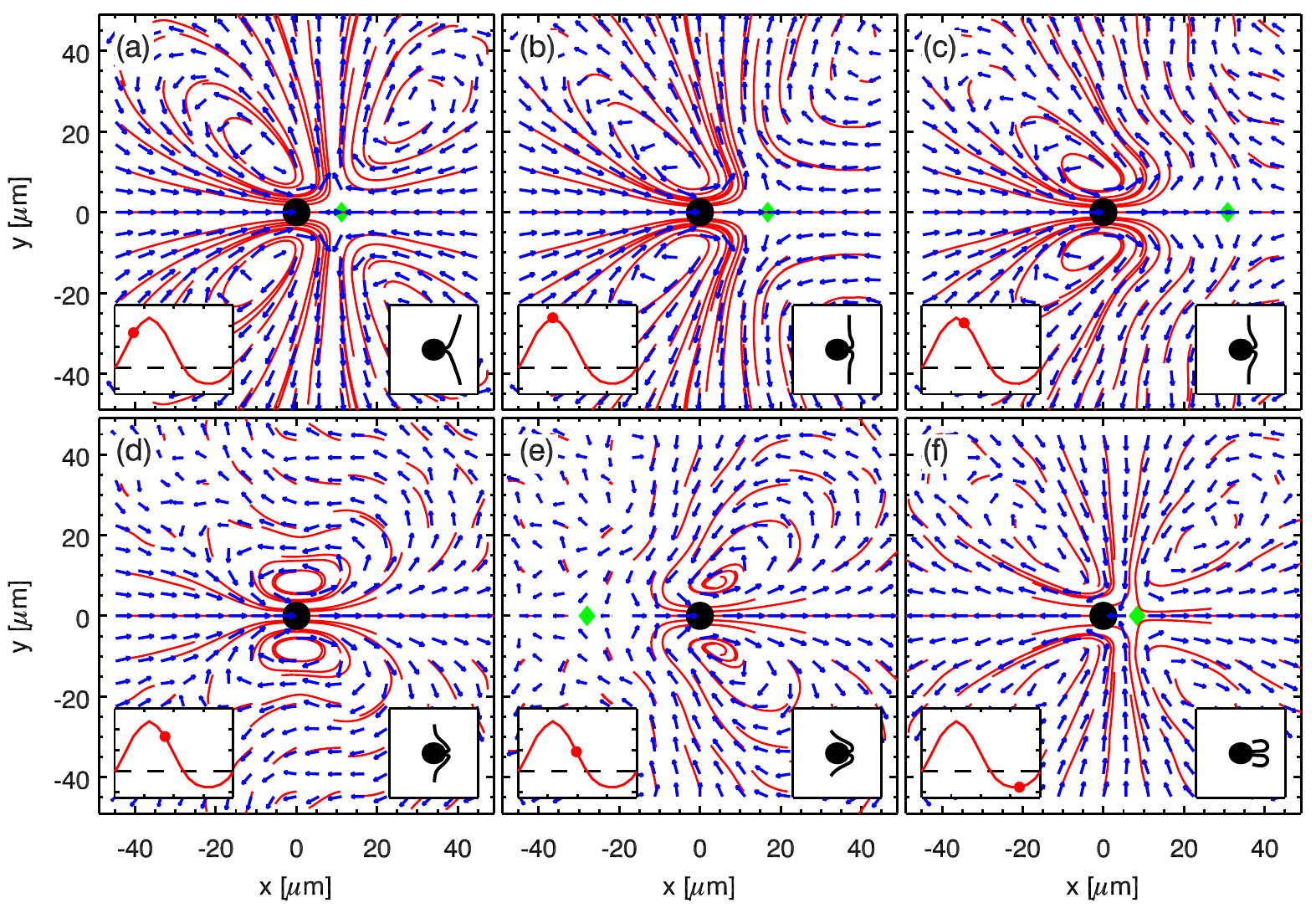}
\end{center}
\caption{(color online) A time sequence of the velocity field evolution throughout the beat cycle of \emph{C. reinhardtii} (period $T=18.9$ ms) along with the position of the hyperbolic stagnation point (green {\Large \textcolor{green}{$\filleddiamond$}}). Insets show cell speed and beat cycle phase (lower left, see Fig. \ref{fig:4}(b) for details), and approximate flagellar shape (lower right) measured from high-speed video. (a) Early in the power stroke, the velocity field resembles a (negative) force dipole. (b) At the peak of the power stroke, the vortices lateral to the organism strengthen and sweep toward the posterior. (c-e) The vortices then shift to the anterior as the power stroke is completed. (f) At the peak of the recovery stroke, the flow field again takes the shape of a dipole, but with opposite sign. The recovery stroke velocity field (f) is weaker than the forward stroke, but is enhanced by the log scaling used in all panels \cite{JSGLogScaling}.} 
\label{fig:3}
\end{figure*}

By simultaneously tracking swimming cells and passive tracer particles (see video \cite{EPAPS}), we measure the velocity field induced by the cells by translating and rotating the instantaneous tracer particle measurements to a common coordinate system based on cellular orientation.
The beat-cycle averaged velocity field is measured at 50 fps, and swimmer trajectory segments with large curvature or irregular beat frequencies are excluded \cite{Polin2009}.
The resulting time-averaged velocity field is shown in Fig. \ref{fig:2}(a), where the swimmer motion is to the right and solid lines are instantaneous streamlines \cite{JSGVelocityField}.
The velocity field is calculated at 3 $\mu$m resolution but represented at 6 $\mu$m resolution using log scaling to show the fine flow features.

While the general shape of the flow field has some qualitative similarities to a force dipole (``stresslet'') approximation \cite{Lauga2009}, several unexpected features are apparent.
The location of the hyperbolic stagnation point far away from the anterior of the cell is surprising, since it typically is thought to be located between the centers of drag (body) and thrust (flagella).
Two strong vortices are visible lateral to the organism (associated with the flagella), while two weaker vortices appear far from the cell body beyond the separatrix of the hyperbolic point.

In the quasi-2D liquid film, we expect a longer-range hydrodynamic disturbance compared to a 3D environment \cite{Prasad2009}.
The magnitude of the fluid velocity normalized by the mean swimmer speed, $u/U_0$, where $u=|\textbf{u}|$, is shown in Fig. \ref{fig:2}(b) as a function of the radial distance from the organism, $r/R$, in various directions.
Toward the posterior, the fluid speed scales nearly as $u \sim r^{-1}$ up to 20 cell radii away, consistent with a force dipole in 2D (compared to $u \sim r^{-2}$ in 3D) \cite{Chwang1975}.
The slight deviation from $u \sim r^{-1}$ scaling is probably due to minor 3D effects.
In the lateral direction, the velocity magnitude passes through a local minimum when traversing the parabolic stagnation points (vortex), before recovering $u \sim r^{-1}$ scaling in the far field.
Similarly, in front of the cell, the fluid speed decreases rapidly near the hyperbolic stagnation point 7-8 radii from the swimmer.

The time-averaged velocity field in Fig. \ref{fig:2}(a) shows the limitations of some current swimmer models, and also raises several important questions. 
For example, how does the velocity field evolve throughout the oscillatory flagellar beat cycle?
Using high-speed imaging (500 fps), we measure the instantaneous swimmer phase, and identify tracer particles at corresponding times in the beat cycle. 
Velocity fields are constructed from tracer velocities at each phase of the oscillation with resolution $T/15$.

A time series of the velocity field during the beat cycle is shown in Fig. \ref{fig:3}, where the cell is again swimming to the right (see video \cite{EPAPS}).
Insets show the swimmer speed and phase (lower left) and the approximate flagellar position (lower right).
At the beginning of the power stroke [Fig. \ref{fig:3}(a)], the velocity field is neatly divided into four symmetric quadrants with the hyperbolic stagnation point located slightly forward from the body. 
This is in line with conventional force dipole swimmer models \cite{Lauga2009}.
As the flagella move toward the posterior and the power stroke peaks, the vortices lateral to the organism strengthen [Fig. \ref{fig:3}(b)] and eventually shift across the body to the anterior side [Fig. \ref{fig:3}(c-e)].
After the power stroke is completed and the recovery stroke begins, the flagella extend out in front of the organism, and the cell velocity becomes negative.
The flow shown in Fig. \ref{fig:3}(f) is qualitatively reversed from Fig. \ref{fig:3}(a), changing the sign of the dipole.
The instantaneous flow field generated by an oscillatory swimmer such as \emph{C. reinhardtii} is complex and highly time dependent.

\begin{figure}[t]
\begin{center}
\includegraphics[clip=true,keepaspectratio,width=7cm]{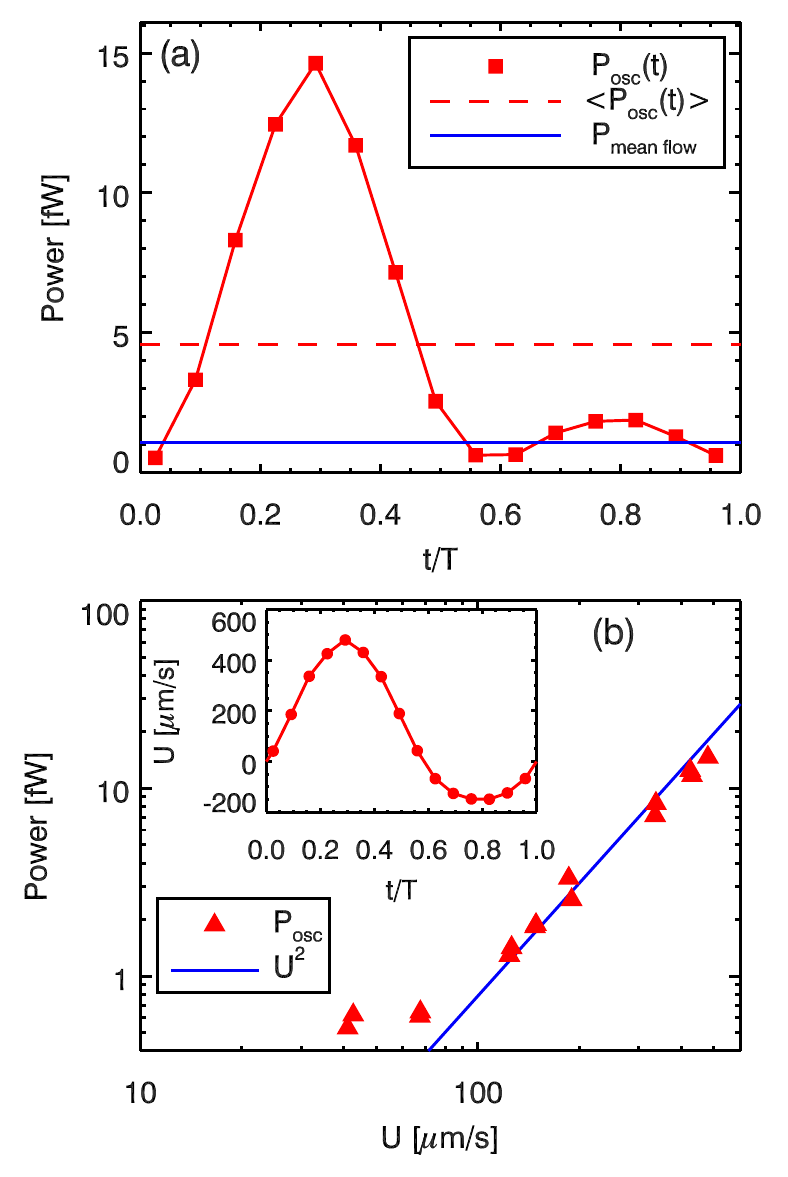}
\end{center}
\caption{(color online) The mechanical power dissipated by the organism is calculated from the velocity fields. (a) The instantaneous dissipation $P_{osc}(t)$ is maximal when the swimmer speed is highest. The dissipated power time-averaged over the beat cycle $\langle P_{osc}(t) \rangle$ is more than four times the dissipation measured from the time-averaged velocity field (i.e. $\langle P_{osc}(t) \rangle/P_{\text{mean flow}} \approx 4.2$). (b) The measured power scales with the square of the cell body speed, $P_{osc} \sim U^2$. Inset: mean cell body velocity as a function of beat cycle phase.}
\label{fig:4}
\end{figure}

In Stokes flows, all of the mechanical energy generated by a swimmer for locomotion is rapidly dissipated by the fluid.
The viscous dissipation per unit volume is given by $\Phi = 2({\bm \Gamma} \vdotdot {\bm \Gamma})$ where ${\bm \Gamma}$ is the rate of strain tensor. 
Thus, the total power transfered to the fluid at any instant by the organism's flagella is $P = \int{\mu \Phi h dA}$, where $\mu$ is the fluid viscosity and $dA$ is a differential area element of the film.
The phase-averaged velocity fields shown in Fig. \ref{fig:3} allow us to calculate the instantaneous mechanical power output,  $P_{osc}(t)$, throughout the oscillatory beat cycle [Fig. \ref{fig:4}(a)].
The peak power output ($\approx 15$ fW) occurs during the power stroke and corresponds to the maximum instantaneous speed of the cell body.
A secondary local maximum also occurs at the peak speed of the recovery stroke [see Fig. \ref{fig:4}(b) (inset)].
Due to the oscillatory swimming, the average mechanical power output over the beat cycle, $\langle P_{osc}(t) \rangle$, is more than four times the power computed from the time-averaged velocity field in Fig. \ref{fig:2}(a) (i.e. $\langle P_{osc}(t) \rangle/P_{\text{mean flow}} \approx 4.2$).

The mechanical power scales with the cell body speed [Fig. \ref{fig:4}(b) (inset)] as $P \sim U^2$ with deviations at small $U$ due to the finite accuracy of measuring tracer velocities [Fig. \ref{fig:4}(b)].
This behavior might be expected for drag-based thrust at low Reynolds number, since the forces scale linearly with the body speed $F \sim U$, and the power output of the organism is $P \sim F U \sim U^2$ \cite{Childress1981}.

This work represents an important step toward understanding the locomotion of swimming microorganisms and collective behaviors for which hydrodynamics are responsible.
Measurements of the time-averaged velocity field around \emph{C. reinhardtii} in quasi-2D thin liquid films demonstrate the predicted $u \sim r^{-1}$ scaling of the fluid velocity with detailed resolution of the near-field flow features.
Also, high-speed imaging allows for phase-resolved measurements of the velocity field throughout the flagellar beat cycle, revealing complex underlying flow structures that evolve in time.
From these measurements, we demonstrate that the mechanical power dissipated by these swimming microorganisms is more than four times what the mean velocity field predicts.
This has important consequences regarding the input of chemical energy required to power oscillatory swimming of flagellated cells.

We thank R.E. Goldstein, K. Drescher, E. Lauga, and M.D. Graham for helpful discussions, as well as B. Boyes for technical assistance. This work was supported by NSF Grant DMR-0803153.


\begin{thebibliography}{}
\bibitem{JSGVelocityField} Velocity fields are averaged about the centerline of the organism and smoothed with a three point average.
\end{thebibliography}
\end{document}